\documentclass[preprint]{aastex631}
\usepackage{amsmath}
\usepackage{float}

\shorttitle{Eccentric orbital planetary engulfment}
\shortauthors{Mengqi et al.}
\graphicspath{{./},{}}

\begin{document}

\title{Engulfment of Eccentric Planets by Giant Stars: Hydrodynamics and Light Curves}

\author[0009-0007-2022-721X]{Mengqi Yang}
\affiliation{Tsung-Dao Lee Institute, Shanghai Jiao Tong University, Shanghai, 201210, China}
\affiliation{Key Laboratory for Laser Plasmas (MoE) and School of Physics and Astronomy, Shanghai Jiao Tong University, Shanghai 200240, China}
\affiliation{Collaborative Innovation Center of IFSA, Shanghai Jiao Tong University, Shanghai 200240, China}
\correspondingauthor{Mengqi Yang}
\email{pachelbel@sjtu.edu.cn}
\correspondingauthor{Dong Lai}
\email{donglai@sjtu.edu.cn}

\author[0000-0002-1934-6250]{Dong Lai}
\affiliation{Tsung-Dao Lee Institute, Shanghai Jiao Tong University, Shanghai, 201210, China}
\affiliation{Center for Astrophysics and Planetary Science, Department of Astronomy, Cornell University, Ithaca, NY 14853, USA}

\author[0000-0003-0482-3365]{Fuyuan Wu}
\affiliation{Key Laboratory for Laser Plasmas (MoE) and School of Physics and Astronomy, Shanghai Jiao Tong University, Shanghai 200240, China}
\affiliation{Collaborative Innovation Center of IFSA, Shanghai Jiao Tong University, Shanghai 200240, China}

\author[0000-0001-7821-4808]{Jie Zhang}
\affiliation{Tsung-Dao Lee Institute, Shanghai Jiao Tong University, Shanghai, 201210, China}
\affiliation{Key Laboratory for Laser Plasmas (MoE) and School of Physics and Astronomy, Shanghai Jiao Tong University, Shanghai 200240, China}
\affiliation{Collaborative Innovation Center of IFSA, Shanghai Jiao Tong University, Shanghai 200240, China}





\begin{abstract}

Recent observations suggest that planetary engulfment by a giant star may produce radiation that resembles subluminous red novae. We present three-dimensional hydrodynamical simulations of the interaction between an eccentric $5 \,M_J$ giant planet and its $1\,M_\odot$ red-giant host star. The planet's pericenter is initially $60\%$ of the stellar radius and is fully engulfed after tens of orbits. Once inside the stellar envelope, the planet generates pressure disturbances that steepen into shocks, ejecting material from the envelope. We use post-processing to calculate the light curves produced by planetary engulfment. We find that the hot stellar ejecta enhances the stellar luminosity by several orders of magnitude. A prolonged hydrogen recombination plateau appears when the ejecta cools to about $10^4\,\rm{K}$. The late-time rapid dimming of the light curve follows dust formation, which obscures the radiation. For planets with lower eccentricity, the orbital decay proceeds more slowly, although the observable properties remain similar.

\end{abstract}



\section{Introduction} \label{sec:intro}

Planets orbiting stars outside our Solar System have been observed over the past few decades, with nearly 6,000 confirmed discoveries\footnote{https://exoplanetarchive.ipac.caltech.edu/index.html} to date. As the host star evolves and its envelope expands, the planets may be ``absorbed" by the stars, a phenomenon known as planetary engulfment~\citep{2009ApJ...705L..81V, 2013ApJ...772..143S,2025ApJ...983...87L}.

Planetary engulfment can produce a variety of observable signatures. The transfer of orbital angular momentum from the engulfed planet to the stellar envelope may change the star’s rotation~\citep{1983ApJ...265..972P, 2018ApJ...864...65Q, 2020ApJ...889...45S}. The deposition of lithium ($^7$Li) in the stellar envelope can also increase its observed abundance~\citep{1998ApJ...506L..65S, 2021AJ....162..273S}. Additionally, the radiation emitted from ejecta gas during engulfment can boost the star’s luminosity, leading to either transient or secular brightness enhancements~\citep{2012MNRAS.425.2778M, 2017MNRAS.468.4399M, 2018ApJ...853L...1M}. A recent observation of an infrared transient, ZTF SLRN-2020, provided evidence for planetary engulfment~\citep{2023Natur.617...55D}.

During the evolution of a planetary system, a planet’s orbit can be excited to high eccentricity through planet–planet scatterings (e.g., see~\citet{1996Sci...274..954R, 1996Natur.384..619W, 1997ApJ...477..781L, 2001Icar..150..303F, 2008ApJ...686..580C, 2008ApJ...686..603J, 2014ApJ...786..101P, 2020MNRAS.491.1369A, 2021MNRAS.501.1621L}, and references therein) or through perturbations from external stellar or planetary companions via the Lidov-Kozai mechanism~\citep{2003ApJ...589..605W, 2007ApJ...669.1298F, 2016MNRAS.456.3671A}. Highly eccentric planets may experience periodic tidal dissipation~\citep{2019MNRAS.484.5645V} and excite stellar mass losses during pericenter passages~\citep{2016ApJ...827...92S, 2019MNRAS.487.3029S}. If the eccentricity continues to grow, a planet can directly enter the stellar envelope before tidal forces circularize its orbit, resulting in engulfment.

A few hydrodynamic simulations of planetary engulfment have been conducted, but none have considered the case of eccentric orbital planets. \cite{1998ApJ...506L..65S, 2002ApJ...572.1012S} used a nested grid to simulate a planet spiraling into the stellar envelope on circular orbit. They found that the dissolution of the planet enhances the surface metallicity. Wind-tunnel simulations~\citep{2015ApJ...803...41M, 2017ApJ...838...56M, 2023ApJ...954..176Y} focused on the local flow surrounding the planet to model the gravitational and hydrodynamic drag forces. Global simulations~\citep{2016MNRAS.458..832S, 2022arXiv221015848L} followed the evolution of the planet's orbit and determined the mass distribution around the planet, as well as the response of the stellar envelope. However, these works did not calculate the luminosity associated with the engulfment, and were limited by the artificial heating of the photosphere in SPH codes~\citep{2016MNRAS.458..832S} or the lack of resolution of the stellar surface in the simulations \citep{2022arXiv221015848L}.

In this study, we conduct $3$D hydrodynamic simulations of the engulfment of eccentricity ($e=0.45,\,0.65$) giant planets with a mass of $5\,M_J$ by a $1\,M_\odot$ evolved red giant star. Note that while close encounters between eccentric giant planets and main-sequence stars can certainly happen, such encounters lead to tidal disruption of the planet near the stellar surface. For giant stars, deep planet engulfment is possible. We adopt the Athena++ code~\citep{2020ApJS..249....4S} for our hydrodynamical simulations and use the outputs to compute the light curve in post-processing. 

The rest of the paper is organized as follows. In Section~\ref{sec:method}, we describe our numerical setup, the method for calculating the light curve and mass loss, and an analytical model for the planetary trajectory. Section~\ref{sec:results} presents our fiducial $3$D simulation results (with the planet eccentricity $e=0.65$), including the hydrodynamic response of the stellar envelope (Section~\ref{subsec:hydro}), planetary orbital migration (Section~\ref{subsec:od}), light curve and photospheric temperature and radius (Section~\ref{subsec:lc}), and mass loss of the ejected gas (Section~\ref{subsec:ml}). Section~\ref{sec:lowe} examines the case of a planet with a lower eccentricity ($e=0.45$), holding other parameters fixed. In Section~\ref{sec:discussion}, we summarize our main conclusions, and discuss the implications and uncertainties.

\section{Numerical Method, setup and analytical model} \label{sec:method}

To investigate the response of the stellar envelope following planetary engulfment, we perform hydrodynamic simulations using the Athena++ code \citep{2020ApJS..249....4S}. Section~\ref{subsec:setup} details the simulation setup for modeling the stellar envelope and the planet. Sections~\ref{subsec:lightcurve} and \ref{subsec:massloss} describe our methods for calculating the light curve and mass loss. In Section~\ref{subsec:analyticalmodel}, we present the analytical model for the planetary trajectory.

\subsection{Numerical setup}\label{subsec:setup}

Our numerical treatment of binary interactions is similar to that described in \cite{2018ApJ...863....5M, 2019ApJ...877...28M, 2020ApJ...893..106M, 2020ApJ...895...29M, 2023NatAs...7.1218M}. We model the dynamical response of the star under the gravitational influence of the planet by solving the Eulerian hydrodynamic equations in spherical coordinates, excising the stellar core. The planet is represented as a gravitational source using a spline function with a softening length of $R_p=0.05\,R_*$, where $R_*$ is the stellar radius. We adopt an ideal gas equation of state (EoS) with an adiabatic index $\gamma = 5/3$, so that the pressure is related to the internal energy density $u$ via $P = 2u/3$.

To obtain a realistic stellar profile, we use the $1$D stellar evolution code MESA~\citep{2011ApJS..192....3P, 2013ApJS..208....4P, 2015ApJS..220...15P, 2018ApJS..234...34P, 2019ApJS..243...10P} to evolve a star of initial mass $M_* = 1\,M_\odot$ to the red-giant phase, reaching a radius of $R_* = 100\,R_\odot$. Our model has a luminosity of $1.22 \times 10^3\,L_\odot$ and an effective temperature of $3406\,\rm{K}$. After obtaining the $1$D profile, we map the density and pressure to the $3$D Athena++ grid, and excise the core within a radius of $R_c=0.3\,R_*$. We use $P = 2u/3$ to compute the internal energy density from the pressure in each cell. The mapped $3$D profile must satisfy hydrostatic equilibrium; if not, we relax the stellar envelope so that it reaches equilibrium. This step is particularly important for early red giant stars. For our MESA giant star model, no envelope adjustment~\citep{2002ApJS..143..539Z} is required, resulting in a similar initial stellar luminosity in Athena++, as shown in Section~\ref{subsec:lc}. For numerical reason, we embed a low-density, constant sound speed ($c_s=\mathrm{constant}$) gas (satisfying $P=\rho c_s^2/\gamma$) outside the star and allow the system to relax for approximately $6\,t_\mathrm{dyn}$ to establish a stable stellar profile, where $t_\mathrm{dyn}$ is the star's dynamical time
\begin{equation}\label{eq:tdyn}
t_{\mathrm{dyn}} = \sqrt{\frac{R_*^3}{G M_*}} = 18.6\,\mathrm{days}.
\end{equation}
The sound speed $c_s$ is chosen to be much larger than the actual photosphere value. Tests conducted with various ambient gas profiles indicate that they have no significant impact on our results.

For our fiducial simulation, we set the planet mass to $M_p = 5\,M_J$ and initially place it at the apocenter of an eccentric orbit with eccentricity $e = 0.65$ and semi-major axis $a = 1.35\,R_*$. After the star is fully relaxed, the planet is released, and the star-planet system is evolved dynamically. Upon engulfment, the planet is expected to eventually dissolve deep within the stellar interior, at a radius smaller than our simulation inner boundary. Therefore, once the planet reaches the inner boundary, we artificially set its mass to zero. The simulation then continues to capture the response of the stellar envelope following the planet’s disruption. The total simulation duration exceeds $300\,t_{\mathrm{dyn}}$.

Our computational domain employs a spherical-polar mesh with resolution $\left(N_r, N_\theta, N_\phi\right) = (192, 96, 192)$. The radial domain extends from $R_c=0.3\,R_*$ to $100\,R_*$ with logarithmically spaced zones. The angular domain covers the full $4\pi$ steradians, spanning $0$ to $\pi$ in polar angle and $0$ to $2\pi$ in azimuthal angle.

\begin{figure}[H]
\centering
\includegraphics[width = \columnwidth]{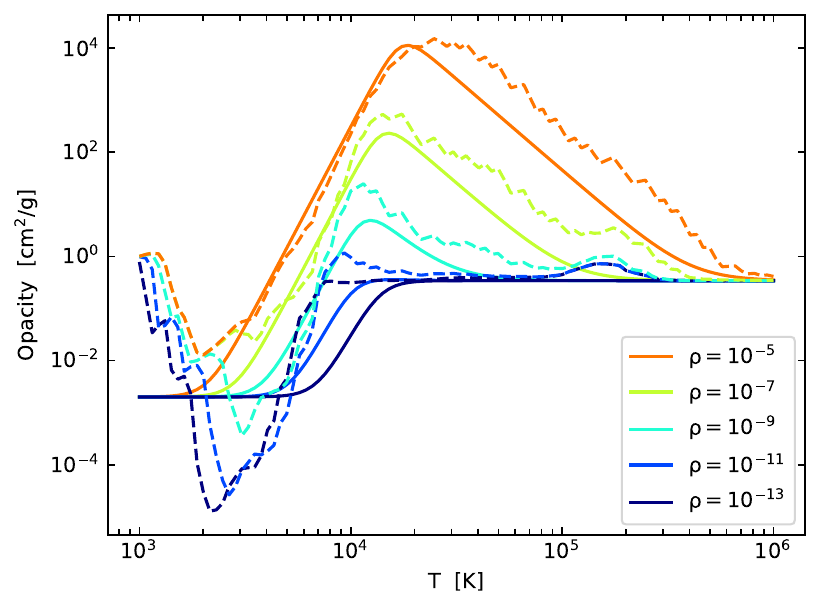}
\caption{Opacities of the analytical model (solid lines) and from the MESA tables (dashed lines) as a function of temperature for selected densities (as labeled, in $\mathrm{g/cm^3}$).
\label{fig:kappa}}
\end{figure}

\subsection{Light curve calculation}\label{subsec:lightcurve}
Planetary engulfment produces the shocked ejecta gas. To compute the luminosity, we first find the position of the photosphere, where the optical depth
\begin{equation}
    \int^{\infty}_{r_\mathrm{pho}} \kappa \rho\, \mathrm{d}r = 1,
\end{equation}
with $\kappa$ the gas opacity and $r_\mathrm{pho}$ the photosphere radius. Since Athena++ does not allow vacuum, we put an artificial ambient gas outside the star, as described in Section~\ref{subsec:setup}. To account for the nonzero optical depth of the ambient gas, we employ a passive scalar $r_*$
 (specifying the concentration of the stellar component in each cell) associated with the stellar material. The optical depth integration only includes gas with a stellar mass fraction exceeding a specified threshold ($r_*\geq 0.9$); tests with other thresholds, such as $0.6$, demonstrate negligible differences in the results.

We use two opacity models: analytical formula and lookup tables from MESA. Each model assumes the hydrogen mass fraction $X=0.7$ and solar metallicity $Z=0.02$. The analytic opacity expression $\kappa$ is from  \cite{2017ApJ...850...59P}:
\begin{equation}
    \kappa = \kappa_m + \left[ \kappa_{H^-} +\left( \kappa_e + \kappa_K  \right)^{-1}   \right]^{-1},
\end{equation}
where $\kappa_m = 0.1Z$ is the molecular opacity, $\kappa_{H^{-}}=1.1 \times 10^{-25} Z^{0.5} \rho^{0.5} T^{1.7}$ is the $H^-$ opacity, $\kappa_e = 0.2(1+X)$ is the electron scattering opacity, and $\kappa_K = 4 \times 10^{25}(1+X) Z \rho T^{-3.5}$ is the Kramers opacity, all in cgs units.

The lookup opacity table is obtained by combining the opacity tables with the file names \textit{gs98\_z0.02\_x0.7.data} and \textit{lowT\_fa05\_gs98\_z0.02\_x0.7.data}~\citep{1998SSRv...85..161G} from MESA. To get the opacity, we use a similar strategy as MESA by linearly interpolating in the log space of density and temperature. Table \textit{gs98\_z0.02\_x0.7.data} is used for high temperature, $\log \mathrm{T}>3.88$, \textit{lowT\_fa05\_gs98\_z0.02\_x0.7.data} for low temperature, $\log \mathrm{T}<3.8$, and ``blending" them in the region $3.8-3.88$. To avoid discontinuities, quintic smoothing is used for blending. Examples of the two opacities as a function of temperature for several chosen densities of interest are shown in Figure~\ref{fig:kappa}.

The two opacity models give similar results with two main differences. First, the MESA opacity model shows significant dust formation below $3000\,\mathrm{K}$, while the analytical formula gives the same low molecular opacity of low temperature ($T\sim 3000\,\mathrm{K}$) for all densities. Second, the MESA opacity model shows a decline at $T\sim(6-8)\times 10^3\,\mathrm{K}$ for low densities (the light blue and dark blue lines in Figure \ref{fig:kappa}), which indicates the onset of the recombination of hydrogen; the analytical formula shows recombination starts around $10^4\,\mathrm{K}$.

Once we find the radius of photosphere, we use the local gas temperature to calculate the specific radiation intensity $I_\nu=B_\nu(T)$, where $B_\nu(T)$ is the Planck function. Since the Athena++ simulation outputs include only the internal energy density and pressure, we apply the ideal gas EoS in post-processing to get temperature
\begin{equation}
    P=\rho k_BT/(\mu m_p),
\end{equation}
where $\mu\simeq 0.625$ is the mean molecular weight of solar abundance and $m_p$ is proton mass (not to be confused with planet's mass $M_p$). 

The specific ``radial" luminosity $L_\mathrm{rad,\,\nu}$ is then obtained by integrating the radiation flux along the radial direction over all solid angles around the source,
\begin{equation}
    L_{\mathrm{rad},\ \nu}  = \int_{0}^{2\pi} \int_{0}^{\pi} \pi B_\nu(T_{\mathrm{pho}}) r_{\mathrm{pho}}^2 \sin\theta  \mathrm{d}\theta  \mathrm{d}\phi, 
\end{equation}
where $T_{\mathrm{pho}}$ is the photospheric temperature. The total ``radial" luminosity $L_\mathrm{rad}=\int \mathrm{d}\nu L_{\mathrm{rad},\ \nu}$ is 
\begin{equation}
    L_{\mathrm{rad}}  = \int_{0}^{2\pi} \int_{0}^{\pi} \sigma_\mathrm{SB} T_{\mathrm{pho}}^4 r_{\mathrm{pho}}^2 \sin\theta  \mathrm{d}\theta  \mathrm{d}\phi, 
\end{equation}
where $\sigma_\mathrm{SB}$ is the Boltzmann-Stefan constant.

We calculate the spherically averaged photosphere radius by
\begin{equation}
    R_\mathrm{pho} = \frac{1}{4\pi}\int r_\mathrm{pho}(\theta , \phi) \sin\theta \mathrm{d}\theta \mathrm{d}\phi.
\end{equation}
Then, the spherically averaged photosphere temperature is given by,
\begin{equation}
    T_\mathrm{pho} = \left[ L_\mathrm{rad}/\left( 4\pi \sigma_\mathrm{SB} R^2_\mathrm{pho}\right) \right]^{\frac14}.
\end{equation}

Since the ejecta produced by planetary engulfment is highly non-spherical, the radial luminosity only represents a crude total energy output from the event. To compute the observed flux, for example along the $x$-direction, $f_x=L_x/(4\pi D^2)$, where $D$ is the distance to the observer, we determine the location of the photosphere in the $x$-direction using
\begin{equation}
    \int^{\infty}_{x_\mathrm{pho}} \kappa \rho\, \mathrm{d}x = 1,
\end{equation}
on a grid in the $yz$ plane (perpendicular to the line-of-sight). Then the ``$x$-luminosities" are given by
\begin{subequations}
\begin{align}
    L_{x,\ \nu}&=4\pi\iint\mathrm{d}y\mathrm{d}z B_\nu(T_{\mathrm{pho,\ x}}),\\
    L_x&=4\iint\mathrm{d}y\mathrm{d}z \sigma_\mathrm{SB} T_\mathrm{pho}^4.
\end{align}
\end{subequations}
where $T_{\mathrm{pho,\ x}}$ (which depends on $y$, $z$) is the $x$-photosphere temperature. The observed luminosities from other directions (e.g., $y$- and $z$- directions) can be obtained similarly.

\subsection{Mass loss calculation}\label{subsec:massloss}
When the planet penetrates the star's envelope, it drives gas ejection through shock. A fraction of the ejected gas is unbound. We compute the cumulative runaway gas at different radius $r$ over time using
\begin{equation}\label{eq4}
    \Delta M_\mathrm{tot} \left( t\right) = \int \mathrm{d}t \int \rho v_r {r^2} \sin\theta \mathrm{d}\theta \mathrm{d}\phi, 
\end{equation}
where $\rho$ and $v_r$ are the gas density and radial velocity. To determine the amount of unbound gas, we evaluate the Bernoulli parameter $B$ of the gas and require it to be positive. Thus
\begin{equation}
\Delta M_\mathrm{unb} \left( t\right) = \int \mathrm{d}t \int_{B>0} \rho v_r r^2 \sin\theta \mathrm{d}\theta \mathrm{d}\phi, 
\end{equation}
The Bernoulli parameter is defined as
\begin{equation}
B =\frac12 \rho v^2 + \rho \phi + u + P,
\end{equation}
where $v$ is gas velocity, $\phi$ is the gravitational potential, and $u = P/\left( \gamma -1\right)$ is the gas internal energy density.

\subsection{Analytical model for planet trajectory}\label{subsec:analyticalmodel}
To compare with our numerical results, we use a semi-analytic model for the planet's motion that includes a simplified gravitational force and dynamical friction:
\begin{equation}\label{eq:force}
    \mathbf{F} = \mathbf{F}_G+\mathbf{F}_g,
\end{equation}
with
\begin{subequations} \label{eq:example}
\begin{align}\label{eq: force}
    \mathbf{F}_G &=-GM_\mathrm{*,\mathrm{enc}}M_p\mathbf{r}/r^3,\\
    \mathbf{F}_g &= C_g \pi R^2_a \rho v\mathbf{v},
\end{align}
\end{subequations}
where $M_{*,\mathrm{enc}}(r)$ is the stellar mass enclosed by the planet's orbit at radius $r$, $\rho$ is the local stellar density and $R_a = 2GM_p/v^2$ is the Bondi-Hoyle radius. The dimensionless coefficient $C_g$ is given by the Coulomb logarithm,
\begin{equation}\label{eq:cg}
    C_g = \ln\left(\frac{S_{\max}}{S_{\min}}\right),
\end{equation}
where $S_{\max}$ and ${S_{\min}}$ correspond to the maximum and minimum impact parameters of the interaction~\citep{1999ApJ...513..252O}. Following \cite{2016A&A...589A..10T}, we use the stand-off distance of the shock $R_\mathrm{sh}$ for $S_{\min}$:
\begin{equation}
    R_\mathrm{sh}=R_p \max(1,\eta),
\end{equation}
where $R_p$ is the planetary radius (or the softening length of the planet), and $\eta$ is a  nonlinearity parameter given by
\begin{equation}\label{eq:eta}
    \eta \equiv \frac{g(\gamma)}{2} \frac{\mathcal{M}^2}{\mathcal{M}^2-1}\frac{R_a}{R_p}.
\end{equation}
We set $\gamma=5/3$ and the corresponding factor $g(\gamma)=1$, and $\mathcal{M}=v_p/c_s$ is the mach number of planet. There are some
uncertainties in the value of $S_{\max}$ for inhomogeneous medium (The \cite{1999ApJ...513..252O} formula applies to homogeneous medium). In general, we can write
\begin{equation}\label{eq: comparison}
S_{\max}=L_\mathrm{int}\left[{\mathcal{M}^2}/{(\mathcal{M}^2-1} )\right]^\frac12,
\end{equation}
where $L_\mathrm{int}$ is the effective interaction distance. For nearly circular trajectory, $L_\mathrm{int}\sim\sqrt{R_*^2-r^2}$ could be adopted. This overestimates the interacting distance in the stratified envelope. \cite{2023ApJ...950..128O} adopted the density scale hight $L_\mathrm{int}\sim H_\rho=\rho/|\mathrm{d}\rho/\mathrm{d}r|$. However, since the bending bow shock is not aligned with the local density gradient, using $L_\mathrm{int}\sim H_\rho$ underestimates $S_{\max}$. An improved estimate for $L_\mathrm{int}$ is to use the density gradient in the direction of $\mathbf{v}$, i.e., $L_\mathrm{int}\sim H_{\rho,\,\hat{v}}=\rho/|{\hat{v}}\cdot\nabla\rho|$· For nearly circlar orbit, we have
\begin{equation}\label{eq: thiswork}
L_\mathrm{int}\sim H_\rho\frac{\left(R_*^2-r^2\right)^\frac12}{R_*-r}.
\end{equation}

Using Eq.~\ref{eq:force}, the planetary trajectory can be integrated using fourth order Runge-Kutta method in the orbital plane, with same initial conditions and stellar profile (non-evolving) as in the simulation. Three different choices of $S_{\max}$ are used for the drag force coefficients. The comparison of the semi-analytical results with the simulation result is shown in Section~\ref{subsec:od}.

\section{results} \label{sec:results}
\subsection{Morphology of the disturbed stellar envelope and ejecta}\label{subsec:hydro}
\begin{figure*}[t]
\centerline{\includegraphics[width = \textwidth]{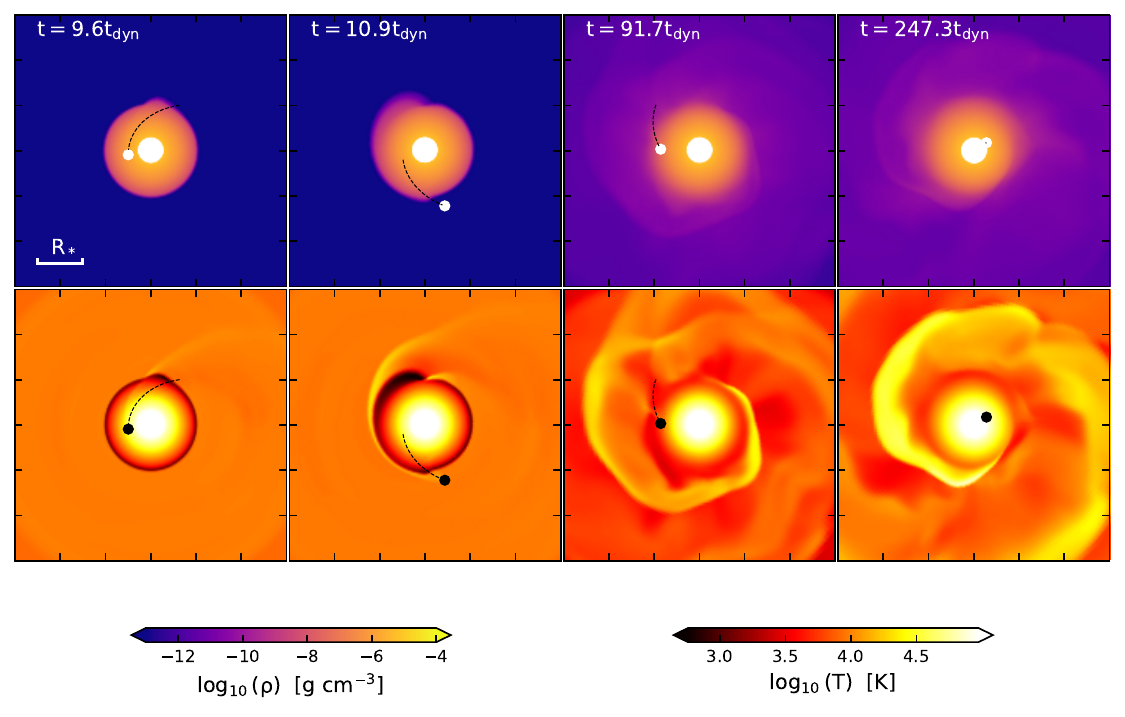}}
\caption{Slices of the density (top row) and temperature (bottom row) through the orbital plane for the fiducial simulation. The left two columns correspond to the first planet–star encounter, followed by snapshots during the planet’s spiral-in and after its disruption. The stellar core is marked by a large white circle, while the planet is indicated by a white (black) circle in the upper (lower) panels. The grid scale corresponds to $1\,R_*$, as shown by the white scale bar in the first panel. The time unit is $t_\mathrm{dyn}$, given by Eq.~\ref{eq:tdyn}.
\label{fig:slice}}
\end{figure*}
\begin{figure*}[ht]
\centerline{\includegraphics[width = \textwidth]{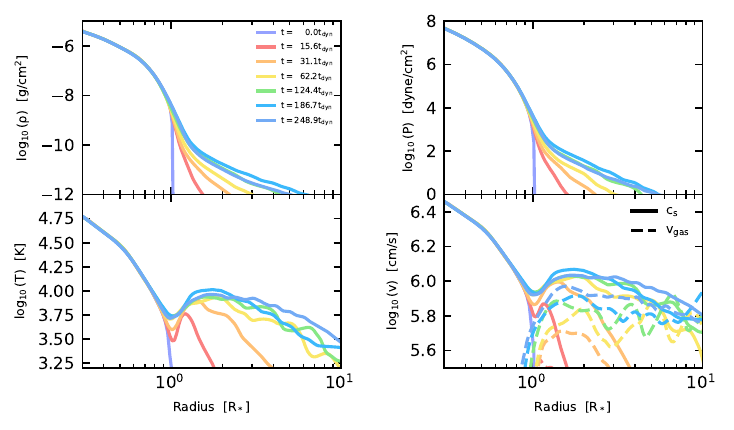}}
\caption{Temporal evolution of the spherically averaged stellar density, pressure, temperature, and sound speed profiles. In the bottom right panel, the spherically averaged gas velocity is shown alongside the sound speed. Different times are indicated by color.
\label{fig:paras}}
\end{figure*}

When the planet directly interacts with the star, the gas exerts a drag force on the planet. Since the star's mean density is quite small compared to that of the planet,
\begin{eqnarray}
        &\overline{\rho_*}/ \overline{\rho_p} &=  \frac{M_*}{M_p} {\left(\frac{R_p}{R_*}\right)}^3  \nonumber\\
        &&\simeq 5.7\times10^{-6}{\left(\frac{M_*}{M_\odot}\right)} {\left(\frac{5M_J}{M_p}\right)} {\left(\frac{R_p}{R_J}\right)}^3 {\left(\frac{100R_\odot}{R_*}\right)}^3,
\end{eqnarray}
the planet comes inside and outside the star several times before the eventual engulfment.

Figure~\ref{fig:slice} shows slices of the stellar gas density and temperature through the orbital plane. After the planet enters the stellar envelope, it drives density disturbances (wakes) via gravity. Due to the planet’s supersonic motion, the wakes steepen into shocks as they propagate. When the wake breaks out of the stellar surface, it ejects stellar envelope gas. At the shock front, kinetic energy is converted into internal energy, producing radiation~\citep{2012MNRAS.425.2778M, 2020ApJ...889...45S}. The typical temperature of the shocked gas is on the order of a few $\times 10^4\,\mathrm{K}$, generating an ultraviolet (UV) transient (see below). After several planet–star encounters, the stellar envelope becomes puffed up, and quasi-periodic spiral-like shocks form once per orbital period (which itself decreases over time).

Figure~\ref{fig:paras} presents the spherically averaged profiles of density, pressure, temperature, and sound speed. As the stellar gas is disturbed, the envelope expands (“puffs up”), resulting in increased density and pressure beyond the original stellar radius, $R_*$. Following the planetary disruption, fallback of ejecta occurs, as indicated by the reduced density and pressure outside the star at $t = 248.9\,t_{\mathrm{dyn}}$ (see the dark blue profile). The gas velocity profiles, shown alongside the sound speed, reveal that the ejecta flow is subsonic at most radii, although the outermost regions exhibit supersonic flow. The peaks in the velocity profile correspond to quasi-periodic shocks within the ejecta.

\subsection{Orbital decay}\label{subsec:od}

\begin{figure*}[ht]
\centerline{\includegraphics[width = \textwidth]{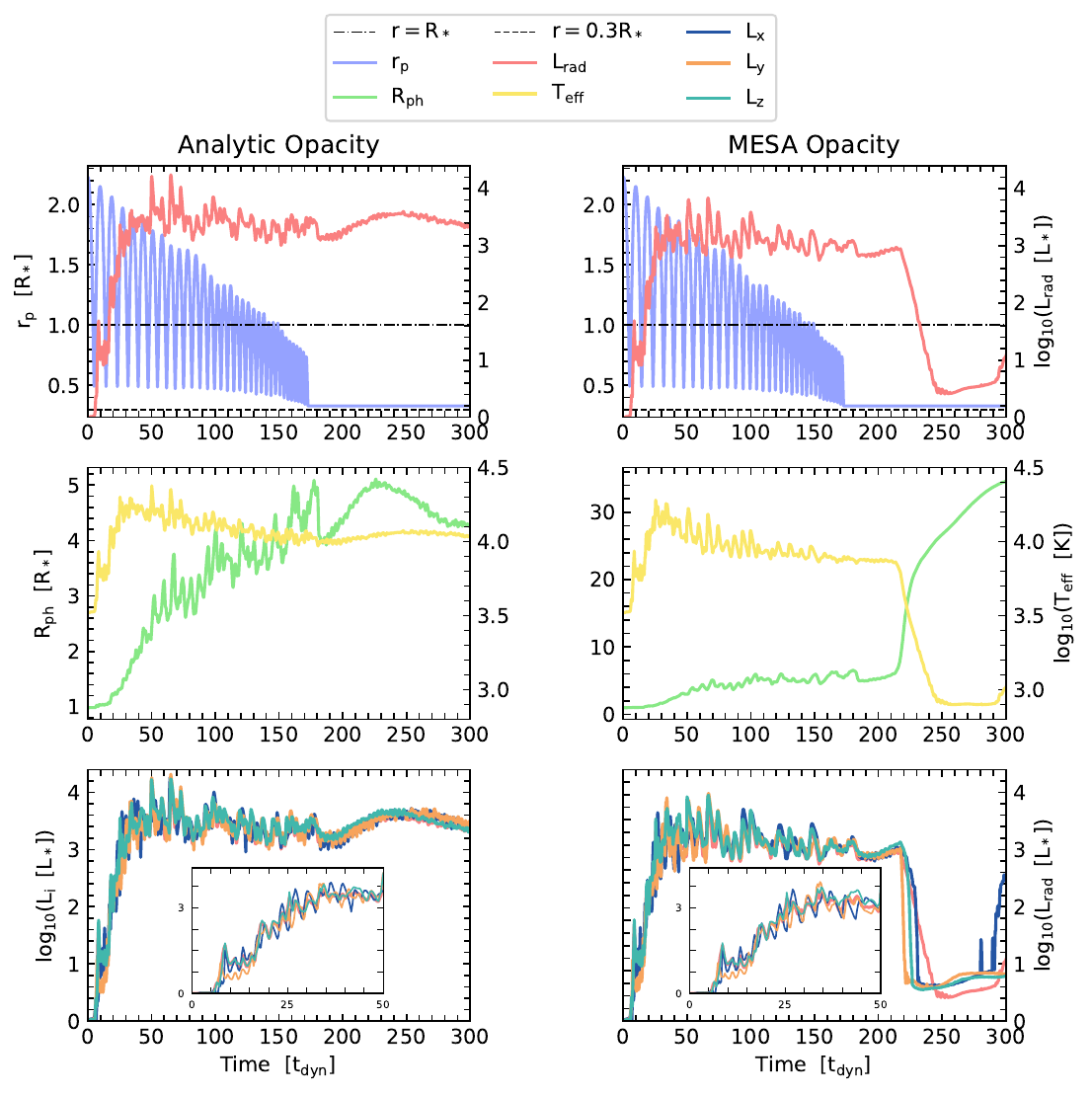}}
\caption{Time evolution of planet's orbital radius $r_p$ (light blue line, top panels), radial luminosity $L_\mathrm{rad}$ (red line, top panels), spherically averaged photospheric radius (light green line, middle panels) and temperature (yellow line, middle panels), observed luminosity (bottom panels) from three different viewing angles ($L_{x}$, dark blue line; $L_y$, brown line; $L_z$, dark green line). The left panels show the observable properties obtained from the analytical formula of opacity, while the right panels are the corresponding results using the MESA opacity tables. In the top panels, the initial star radius and inner boundary radius are indicated by the black dotted and dashed lines, respectively. The radial and observed luminosities are normalized to the initial star luminosity $L_*$. After the planet is disrupted when it reaches the inner boundary at $\sim 175\, t_\mathrm{dyn}$, the orbital radius is set to the radius of the inner boundary ($0.3\, R_*$). 
\label{fig:lc}}
\end{figure*}

\begin{figure}[t]
\centering
\includegraphics[width = 0.8\columnwidth]{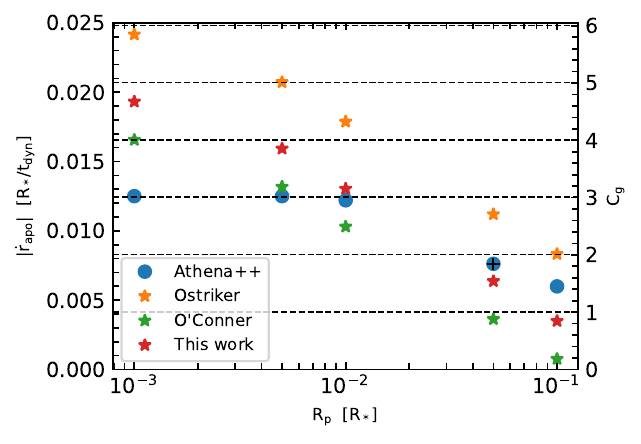}
\caption{The orbital decay rate, $|\dot{r}_{\mathrm{apo}}|$, as a function of the planetary radius $R_p$ (which is the softening radius). The blue circles show the results from Athena++ simulations differing only in the softening length. The stars mark the predictions from the semi-analytical model based on three choices of $L_\mathrm{int}$: orange for $\sqrt{R_*^2-r^2}$ (Ostriker), green for $H_\rho$ (O'Conner), and red for $H_{\rho,\,\hat{v}}$ (this work). The corresponding drag-force coefficient, $C_g$, is shown on the right axis. The fiducial case is indicated by an overlaid black cross.
\label{fig:fig_rdot}}
\end{figure}

The temporal evolution of the planet’s orbital distance from the stellar center is shown in Figure \ref{fig:lc} (blue line, top panels). Before the planet fully descends onto the stellar surface, the apocenter distance decreases quasi-linearly, while the pericenter distance remains nearly constant. We therefore characterize the orbital decay rate by the rate of decrease of the apocenter distance, $|\dot{r}_\mathrm{apo}|$. In the fiducial model, this rate is $\sim 0.012\,R_*/t_\mathrm{dyn}$. Once the planet is entirely submerged beneath the stellar surface, the pericenter distance begins to decline as the drag force increases sharply at greater depths within the envelope. After several orbits, the planet reaches the inner edge of the simulation domain (at $0.3\,R_*$, dashed line in Figure \ref{fig:lc}). At this point, we artificially set the planet’s mass to zero to halt its motion, resulting in a constant orbital distance thereafter.

Recall that in our fiducial simulation, we treat the planet as a gravity source with softening length $R_p=0.05\,R_*$. To test the robustness of our result, we have performed additional simulations with different values of $R_p$, the results are shown in Figure~\ref {fig:fig_rdot} along with the semi-analytic results. Even though the results from Athena++ simulations display a cut-off when $ R_p $ matches the smallest grid size $ \Delta r=0.01\,R_* $, all other orbital decay rates exhibit a linear decrease in relation to the increasing $ R_p $.

As detailed in Section~\ref{subsec:analyticalmodel}, we integrate orbital trajectories under varying drag force coefficients. For a constant coefficient, the orbital decay rate exhibits a linear dependence on the coefficient; for example, $|\dot{r}_\mathrm{apo}|\propto 0.4\,C_g$, as indicated on the right axis. The outcomes for the three $L_{\rm{int}}$ values are represented by stars of different colors. The results derived from Eq.~\ref{eq: thiswork} (red stars) demonstrate the closest alignment with hydrodynamic simulations, suggesting that the optimal estimate for $L_{\rm{int}}$ corresponds to the density scale in the perpendicular direction. The other two selections yield either over- or underestimates of orbital behavior within the stratified medium.

\subsection{Light curve}\label{subsec:lc}

\begin{figure*}[t]
\centerline{\includegraphics[width = \columnwidth]{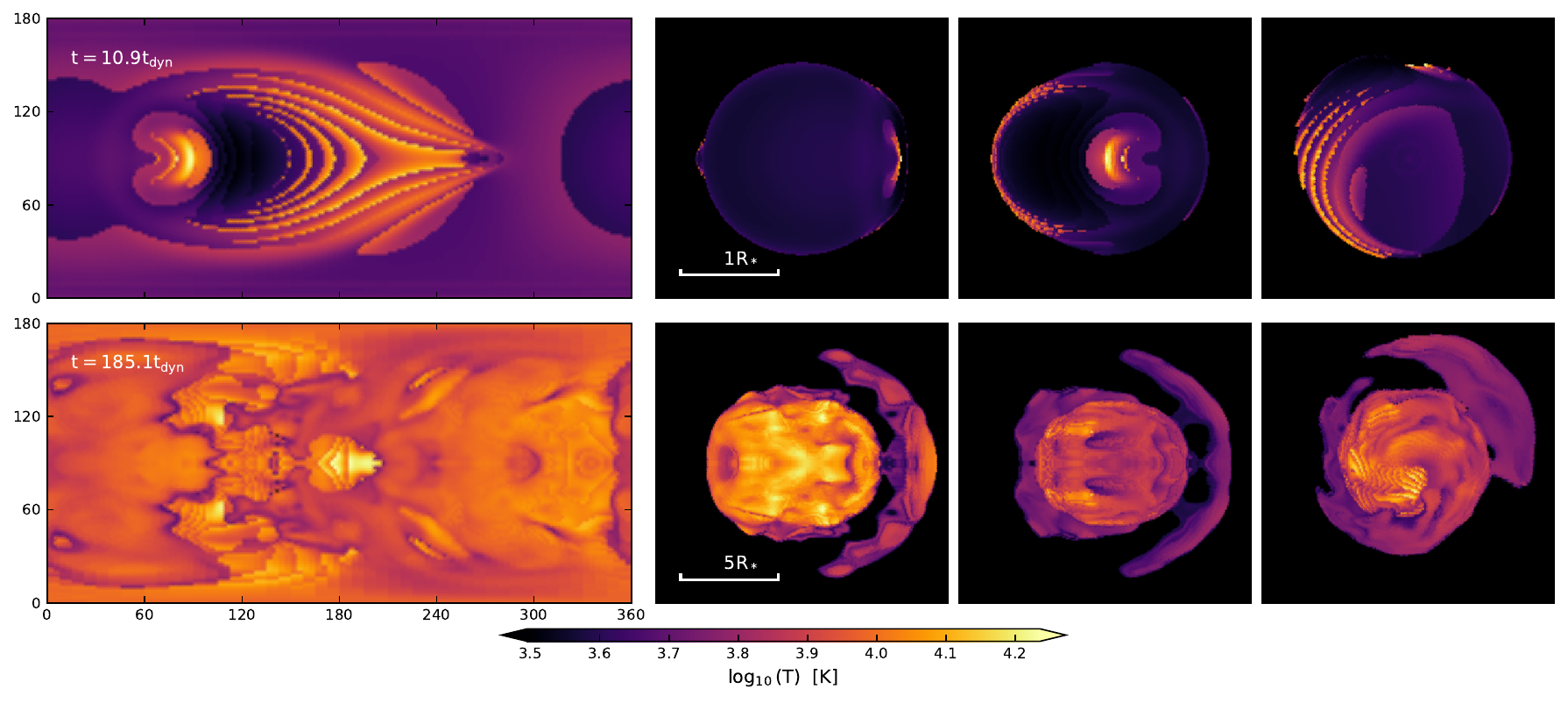}}
\caption{Synthetic photospheric temperature maps for different viewing directions at two selected times. These are the log–log plots of radiation intensity, since for blackbody emission, $\log_{10} I \propto 4 \log_{10} T$. The upper panels correspond to $t=10.9\,t_\mathrm{dyn}$, while the bottom panels correspond to $t=185.1\,t_\mathrm{dyn}$. The leftmost column displays the photospheric temperature in spherical polar coordinates $(\theta, \phi)$, while the right three columns show temperature distributions projected onto the Cartesian planes corresponding to viewing directions along the $+x$, $+y$, and $+z$ axes. The planet's orbital plane is the $xy$ plane.
\label{fig:pho_temp}}
\end{figure*}

We show the spherically averaged radial luminosity, photospheric radius, and effective temperature obtained using both the analytic opacity formula and the MESA opacity tables in Fig.\ref{fig:lc}. The photospheric properties are calculated at every hydrodynamics output, with a time interval of about $0.3\,t_{\mathrm{dyn}}$. As a check for our method, we compare the computed bolometric luminosity and photospheric temperature prior to the planet–star interaction, when the host star is the same as the $1$D result from MESA. The luminosity and photospheric temperature are $L_* = 1.06\times10^3\,L_\odot$ and $T_{\mathrm{eff}} = 3319\,\mathrm{K}$, respectively, differing slightly from the MESA values (see Section~\ref{subsec:setup}) due to the resolution difference between the two grids and the relaxation procedure. After the planet interacts with the host star, the photospheric properties exhibit similar evolutionary stages: onset\footnote{We do not use the term “plunge-in”, which is commonly adopted in the literature. For circular planetary engulfment or common envelope evolution with negligible secondary eccentricity, “plunge-in” refers to the initial phase of the rapid orbital decay. A reasonable indicator of this phase is when the ratio of the orbital period to the orbital decay timescale lies in the range of $0.01 \lesssim P_{\mathrm{orb}}/(a/|\dot{a}|) \lesssim 0.1$~\citep{2020cee..book.....I}. In the eccentric case, this occurs only after the secondary is fully embedded within the stellar envelope.}, hot peak, and plateau. The light curve obtained with the MESA opacity tables additionally shows a late-time dimming phase due to dust formation.

During the onset stage, the planet penetrates the stellar envelope, driving shocks through gravity. These shocks eject gas and enhance the stellar luminosity. The shocked ejecta exhibit a characteristic temperature of approximately $2 \times 10^4\,\mathrm{K}$ (first row of Fig.~\ref{fig:pho_temp}), consistent with the strong-shock assumption from the Rankine–Hugoniot jump conditions:
\begin{equation}\label{eq:t_shock}
T_{\mathrm{shock}} \simeq \frac{3}{16} \mu m_{p} v_{p,\mathrm{out}}^2/k_{B},
\end{equation}
where $v_{p,\mathrm{out}} = \sqrt{GM_*\left({2}/{R_*} - {1}/{a}\right)}\sim 4 \times 10^6\,\mathrm{cm\,s}^{-1}$ is the planet’s velocity upon coming outside the stellar surface. During this onset phase, as the shock wave propagates through the stellar envelope, the spherically averaged photospheric temperature gradually increases while the photospheric radius remains close to the initial stellar radius.

The hot peak stage begins after the planet has encountered the star approximately three to four times, during which the luminosity rises to a few times $10^3\, L_*$. As the shocked ejecta expands, the photospheric temperature decreases due to adiabatic cooling. A notable feature of this stage is the quasi-periodic hot peaks, originating from shock formation during successive planet–star encounters. The shocked gas is ejected as the planet spirals inward, eventually surpassing the expanding photosphere. As the spiral-arm-shaped shocked ejecta cools and becomes optically thin, the underlying inner quasi-spherical, adiabatically expanding gas becomes visible, as illustrated in the bottom row of Figure~\ref{fig:pho_temp}.

The plateau stage begins as a result of hydrogen recombination, a hallmark feature in the light curves of common-envelope events (CEEs) \citep{2013Sci...339..433I}. In models using the MESA opacity, the photospheric temperature during the plateau is lower, approximately $6000\,\rm{K}$, reflecting the fact that recombination occurs at lower temperatures in low-density regions (see Section~\ref{subsec:lightcurve}), whereas models with analytical opacity show a photospheric temperature around $10^4\,\rm{K}$. During this stage, the photosphere begins to recede. The quasi-periodic hot peaks disappear, when the planet is fully disrupted at $t\simeq 170\,t_{\mathrm{dyn}}$.

The dimming stage begins with dust formation in the more distant, cooling gas. Dust forms below temperatures of approximately $1000\,\rm{K}$, increasing the opacity to about $1\,\rm{cm}^2\,\rm{g}^{-1}$ and thereby obscuring thermal emission from the hot ejecta \citep{2025ApJ...982...83H}. We propose that shocks driven by the spiraling-in planet may destroy dust grains in the low-temperature gas, leading to a more gradual dimming of the light curve.

Since the mass ejection is highly nonspherical, the observed radiation depends on the viewing angle. Figure~\ref{fig:pho_temp} presents three-view projections of the photospheric temperature alongside an equirectangular projection of the radial line-of-sight photosphere temperature. Shortly after the first interaction at $t = 10.9\,t_{\mathrm{dyn}}$ (upper panels), the wake shock appears as a spiral-shaped high-temperature region propagating along the equatorial plane following the planet. Much later, as the ejecta expands quasi-spherically at $t = 185.1\,t_{\mathrm{dyn}}$ (lower panels), a faster-propagating shock surpasses the expanded photosphere. Once this shock becomes optically thin, the spherically expanding photosphere emerges, giving rise to luminosity modulation.

The light curves from three observing directions are shown in the bottom panels of Figure~\ref{fig:lc}. The observer lines of sight correspond to the $+x$, $+y$, and $+z$ directions, respectively. The luminosities are normalized by $L_*$ to enable direct comparison with the radially averaged luminosities, which are also plotted in the same panels as red lines. The observed luminosities from different directions generally agree with the radial luminosity for both opacity models. Three-dimensional effects are most pronounced during shock propagation, causing peak delays especially in $L_x$ and $L_y$. The luminosity along the $+z$ direction, $L_z$, is lower than the other two for the first few orbits, primarily due to the smaller projected area of the shocked region.

\subsection{Mass loss}\label{subsec:ml}

\begin{figure*}[ht]
    \centering
    
    \begin{minipage}{0.45\textwidth}
        \includegraphics[width=\textwidth]{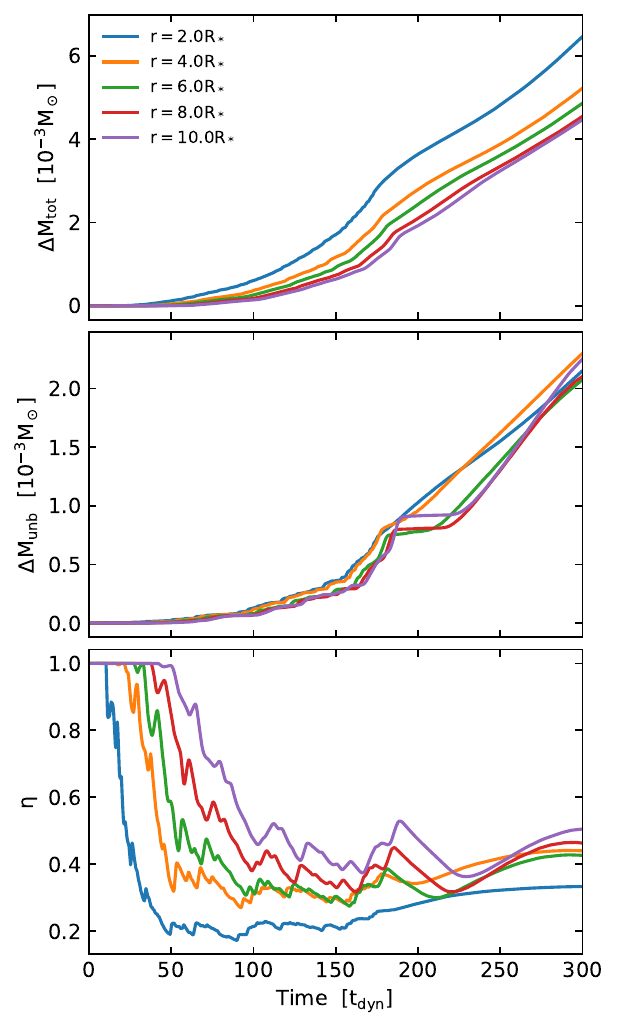}

    \end{minipage}
    \hfill
    \begin{minipage}{0.45\textwidth}
        \includegraphics[width=\textwidth]{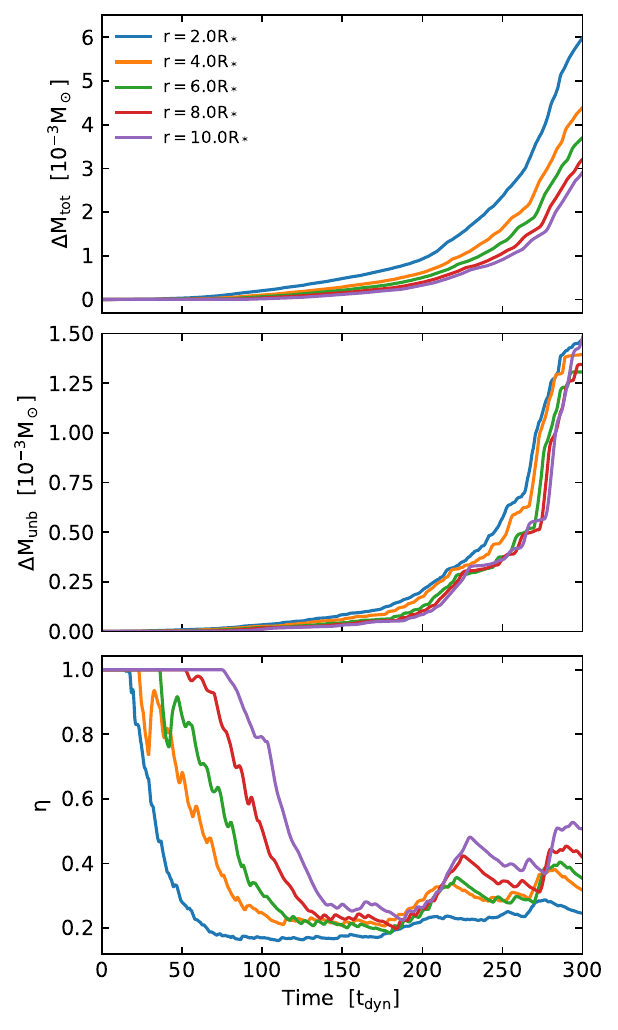}

    \end{minipage}
    \caption{Temporal evolution of the cumulative ejected mass, $\Delta M_{\mathrm{tot}}$ (top panel), the cumulative unbound mass, $\Delta M_{\mathrm{unb}}$ (middle panel), and the unbound mass fraction, $\eta$ (bottom panel), evaluated at various reference radii. The left panels correspond to the fiducial simulation with the initial planet orbital eccentricity $e = 0.65$, while the right panels show the case with $e = 0.45$.
    \label{fig:mass_loss}}
\end{figure*}

The left panels of Figure~\ref{fig:mass_loss} show the cumulative ejected gas mass, unbound mass, and unbound fraction at different reference radii. The planet’s orbital energy deposited into the stellar envelope ejects gas from the gravitational potential well. Gas with a positive Bernoulli parameter tends to stream freely to infinity if unimpeded by further collisions, while gas with negative Bernoulli gathers in the circumstellar environment around the star. 

The total ejected mass grows exponentially to approximately $(2$–$3) \times 10^{-3}\,M_\odot$ before the planet disrupts, then increases roughly linearly to $(4$–$6) \times 10^{-3}\,M_\odot$ at a nearly constant mass-loss rate thereafter. The unbound mass exhibits a similar behavior, with a “step” feature at radii larger than $4\,R_*$ shortly after planet disruption. The unbound mass fraction, $\eta = \Delta M_\mathrm{unb}/\Delta M_\mathrm{tot}$, rapidly declines before settling to an approximately constant value after $50\,t_\mathrm{dyn}$. The higher unbound mass fraction observed at larger radii suggests that a portion of the unbound gas decelerates or stalls at smaller radii.

We note that the mass of the ejected gas continuous to increase at the end of our simulation due to the computation time constraint. For example, the amount of total ejected mass $\Delta M_\mathrm{tot}$ grows at a rate of $\sim1\times10^{-5}\, M_\odot/t_\mathrm{dyn}$ for all the reference radii. The unbound mass $\Delta M_\mathrm{unb}$ grows at a rate of $ \sim4\times10^{-6}\,M_\odot/t_\mathrm{dyn}$, with decreasing fraction of unbound mass for larger radii. Thus, we can only place a lower limit on the full amount of the ejected mass in our simulation.

\section{Lower eccentricity case}\label{sec:lowe}
\begin{figure*}[ht]
\centerline{\includegraphics[width = \columnwidth]{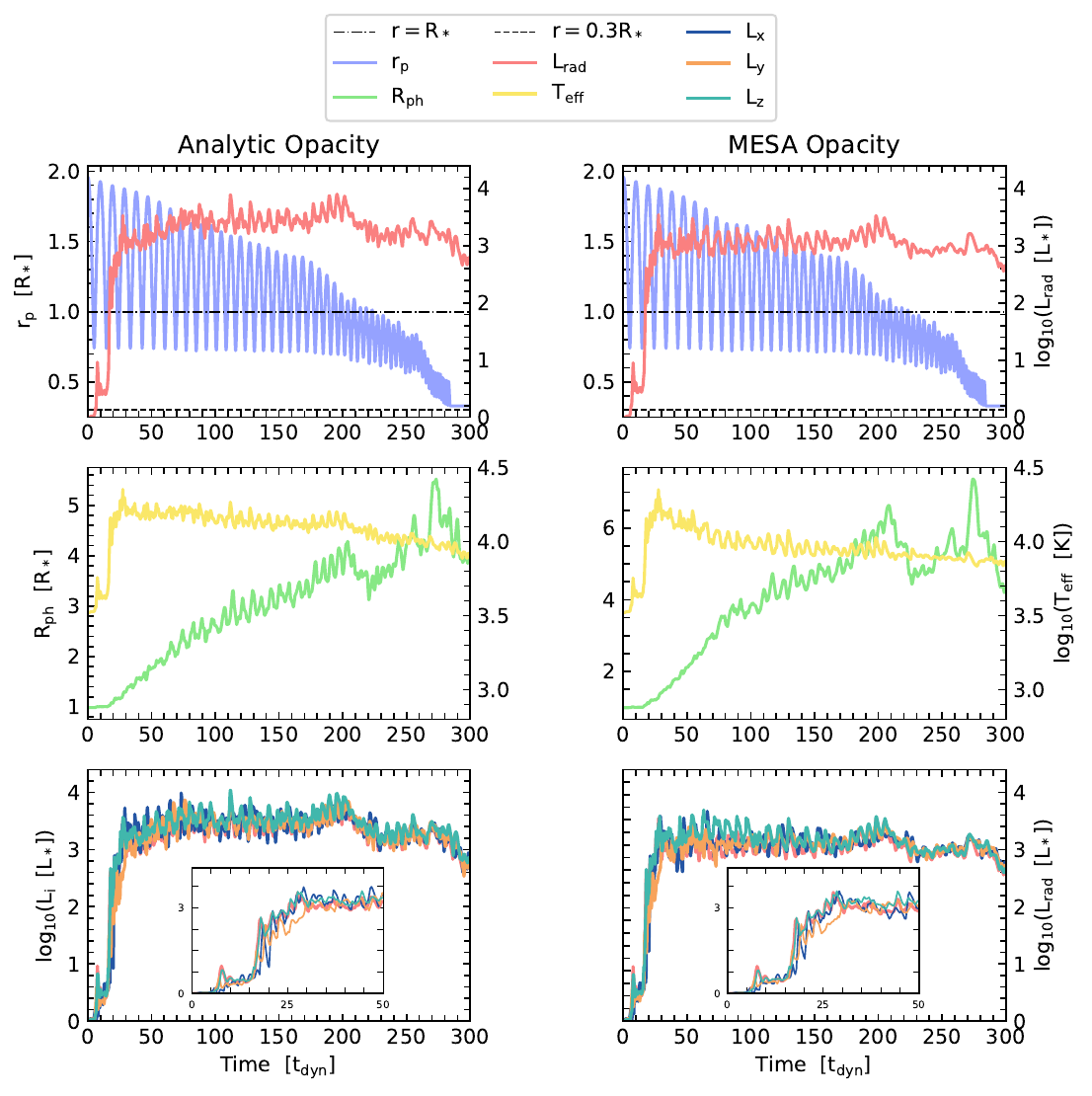}}
\caption{Same as Figure \ref{fig:lc}, except for the case of $e=0.45$ (the initial planet eccentricity).
\label{fig:lc_le}}
\end{figure*}

To investigate the impact of planetary eccentricity, we perform a simulation with $e=0.45$ while keeping all other parameters fixed. The corresponding apocenter and ballistic pericenter distances are $r_\mathrm{apo} \simeq 2\, R_*$ and $r_\mathrm{peri} \simeq 0.74\, R_*$, respectively. At this $r_\mathrm{peri}$, the lower stellar envelope density results in a weaker dynamical friction. Consequently, the orbital decay deposits energy into the stellar envelope more gradually.

We show the orbital decay and observable properties in Figure~\ref{fig:lc_le}. The orbital decay is slower than in the fiducial case ($e=0.65$); the planet takes over $200\,t_{\mathrm{dyn}}$ to fully descend beneath the stellar surface and more than $280\,t_{\mathrm{dyn}}$ to be disrupted. The orbital decay rate defined in Section~\ref{subsec:od} is approximately $0.003\,R_*/t_{\mathrm{dyn}}$, about four times slower.

The light curve exhibits similar features to the fiducial case. The onset phase lasts roughly three orbits before reaching the first peak, with an average photospheric temperature near $2 \times 10^4\,\mathrm{K}$, comparable to the fiducial model. From Eq.(\ref{eq:t_shock}), the shock temperature associated with the planet’s velocity coming out from the stellar surface, $v_{p,\mathrm{out}} = \sqrt{GM_*\left({2}/{R_*} - {1}/{a}\right)}$, is independent of the eccentricity. Another notable difference is that the quasi-periodic peaks in luminosity and photospheric radius seen in the fiducial case are weaker here, possibly because the lower-energy shocks dissipate more rapidly. With the MESA opacity tables, we do not observe the rapid dimming of the light curve, suggesting that dust formation in the hot ejecta is suppressed within our simulation time, as the planet disruption occurs shortly before our simulation ends.

The mass loss is shown in the right panel of Figure~\ref{fig:mass_loss}. Before the planet fully plunges beneath the stellar surface, the mass-loss rate is slower than in the fiducial case, reflecting the more gradual deposition of orbital energy into the stellar envelope. After the planet is completely embedded within the envelope, the orbital radius decreases rapidly in a manner similar to the plunge-in phase \citep{2016ApJ...816L...9O, 2020cee..book.....I}, during which mass loss is enhanced. By the end of the simulation, the cumulative ejected mass is comparable to that of the fiducial case at small reference radii; however, the fraction of unbound mass is smaller, with the unbound ejecta mass remaining below $1.5 \times 10^{-3}\, M_{\odot}$.

\section{Summary and Discussion}\label{sec:discussion}

We have performed three-dimensional ($3$D) hydrodynamics simulations of an eccentric planet interacting with and eventually being engulfed by a red giant host star, using the Athena ++ code. The light curve and mass loss are computed in post-processing. The local gas temperature at the photosphere (optical depth $\tau=1$) is used to estimate the thermal emission. Mass loss is calculated by integrating the mass flux through spherical surfaces at various radii outside the star. Our results suggest that such planetary engulfment can produce transients resembling luminous red novae (LRNe), one of which has been previously identified as the engulfment of a giant planet in a circular orbit by its main-sequence host star \citep{2023Natur.617...55D, 2025ApJ...983...87L}.

Our main findings are:
\begin{itemize}
    \item The direct interaction of planet with its host giant star after a sudden eccentricity excitation can lead to the engulfment of planet.
    \item During the multiple planet-star encounters, the disturbances produced by the planet's gravity steepen into shocks and eject the gas of the stellar envelope. The thermal emission of the gas brightens the stellar luminosity by serval orders of magnitudes.
    \item The light curves exhibit similar characteristics as luminous red novae (LRNe). During the initial engulfment, the stellar luminosity rapidly increases within a few orbits. During the ``hot-peak" stage, the luminosity gradually decreases as the photosphere expands. A long plateau phase follows with nearly constant temperature when the hydrogen recombination happens. The rapid dimming phase occurs as dusts form at temperatures below $\sim 1000\,\rm{K}$.
    \item Due to the quasi-periodic planet-star interactions with the planet's elliptical orbit, wake shocks eject the stellar envelope in quasi-periodic manner, leading to quasi-periodic peaks in the luminosity and photospheric radius.
    \item The ejected mass is roughly a few $\times 10^{-3}\, M_{\odot}$ for a $5\,M_J$ planet engulfed by $1\,M_\odot$ red-giant star from our simulation. Some of the gas stay in the circum-stellar environment, and less than $50\%$ of the ejected mass become unbound.
    \item By comparing with $3$D hydrodynamics simulation, we find a reasonable expression of the coefficient of the gravitational dynamical friction force for the stratified star.
\end{itemize}

The primary limitation of our simulations is the omission of radiation transport. For a low-mass red giant, the thermal timescale is many orders of magnitude longer than the orbital decay timescale, making radiation transport within the stellar interior negligible. Within the ejecta, however, the local radiation diffusion timescale can be short, especially where the optical depth is less than unity. In this regime, the thermal energy carried by radiation is non-negligible, and radiation transport could modify the ejecta temperature, thereby affecting the predicted observational signatures. 

During the ejecta expansion, radiative loss can reduce the thermal energy, which in return decreases the mass-loss rate. Consequently, the sustained mass loss predicted in our calculation—similar to that found in previous work \citep{2025arXiv250320506L}—may be suppressed. However, radiation transport is unlikely to have a significant dynamical impact on the orbital decay \citep{2025arXiv250320506L}.

Although hydrogen recombination appears as a long plateau in the light curve, it is not included in the equation of state (EoS) used in the hydrodynamics simulations or the post-processing temperature calculations. \citet{2024ApJ...963L..35C} studied common-envelope evolution by comparing different EoS and found that neglecting realistic EoS and radiation forces leads to faster light-curve dimming and luminosities reduced by up to a factor of a few. \citet{2022MNRAS.516.4669L} also found that including hydrogen and helium recombination in EoS leads to tens of percent more mass loss. Therefore, our simulation might underestimate the mass loss.

Dust formation plays a key role in shaping the observational signatures of planetary engulfment events, as it can produce a prolonged infrared emission phase \citep{2023Natur.617...55D, 2025ApJ...982...83H}. As explored in this work using the MESA opacity tables, dust formation in the cooling ejecta can obscure the radiation from the hot material, resulting in a rapid late-time decline in the light curve. These dusts, however, may be destroyed by the quasi-periodic shocked ejecta. More detailed modeling of dust formation and destruction is required when applying theoretical predictions to interpret real observations.

In our simulations, we artificially disrupt the planet once it reaches the inner boundary at $0.3\,R_*$. However, the planet could potentially survive much longer. When the planet’s internal temperature rises such that the sound speed exceeds its surface escape velocity, dissolution begins \citep{1984MNRAS.208..763L}. However, considering the limited heat transport via conduction and radiation, which leads to heating timescales much longer than the predicted in-spiral time \citep{1996ApJ...459L..35G, 2023ApJ...950..128O}, the planet is expected to survive within the stellar envelope until it fills its Roche lobe, at which point it undergoes rapid tidal disruption \citep{1999ApJ...524..952R}.

During engulfment, the planet can drive differential rotation and turbulence in the stellar envelope, conditions favorable for magnetic field amplification via a dynamo mechanism. A few MHD simulations have investigated magnetic field amplification during the common-envelope phase \citep{2016MNRAS.462L.121O, 2022A&A...660L...8O, 2024A&A...683A...4G}, consistently finding strong amplification. However, since no simulations have addressed the case of a low-mass companion, the efficiency of magnetic field amplification during planetary engulfment remains uncertain. If the stellar magnetic field is indeed amplified, it could drive a magnetically powered stellar wind \citep{1998MNRAS.299.1242S}. Nonetheless, \citet{2016MNRAS.462L.121O} found that in the common-envelope phase the dynamical impact of the magnetic field is modest, with mass loss enhanced by only $5$–$6\%$ during the first orbit. 

The initial motivation of this study is to understand observable features of the ZTF SLRN-2020 event, interpreted as the engulfment of a giant planet by a red giant in circular orbit. Because of the large eccentricity, our modeled system exhibits a greater luminosity, driven by quasi-periodic shocks and mass ejections. In a circular orbit scenario, envelope ejection is expected during surface interactions—provided the thermal timescale is sufficiently short and the interaction is dynamically significant. In contrast, an eccentric orbit facilitates enhanced mass loss, and yields brighter emission and prolonged UV–optical radiation.

\section*{Acknowledgements}
We thank Mike Y. M. Lau for useful discussions. The simulations were performed on the Siyuan high performance computers at Shanghai Jiao Tong Universities. This work was supported by the National Natural Science Foundation of China (No. 12205185).


\software{Athena++ \citep{2020ApJS..249....4S}, MESA \citep{2011ApJS..192....3P, 2013ApJS..208....4P, 2015ApJS..220...15P, 2018ApJS..234...34P, 2019ApJS..243...10P}, ipython/jupyter \citep{2007CSE.....9c..21P}, NumPy\citep{2023EGUGA..25.8161S}, matplotlib\citep{4160265}, SciPy\citep{virtanen2020scipy}}



\bibliography{PEGM}{}
\bibliographystyle{aasjournal}



\end{document}